\input{aipcheck}

\documentclass[
    ,final            
  ]
  {aipproc}

\layoutstyle{8x11double}
\usepackage{amsmath,amssymb}

\begin{document}

\title{Zero modes, Instantons, and Monopoles}


\author{Adriano Di Giacomo}{
  ,address={University of Pisa, Department of Physics, Largo B. Pontecorvo, 3, PISA, 56127, ITALY}
}

\author{Masayasu Hasegawa}{
  address={Bogoliubov Laboratory of Theoretical Physics, Joint Institute for Nuclear Research, Dubna, Moscow Region, 141980, RUSSIAN Federation}
  ,altaddress={Previous affiliation: University of Parma and INFN, Department of Physics, Via G. P. Usberti 7/A, Parma, 43124, ITALY}
}

\begin{abstract}
The purpose of this study is to show the relations between monopoles instantons and Chiral symmetry breaking. First, in order to show the relation between instantons and monopoles, we generate configurations, adding monopoles by a monopole creation operator. Then, we count the number of fermion zero modes in the configurations using Overlap fermions as a tool. As a result we find that one monopole with plus one charge and one anti-monopole with minus one charge make one instanton of charge plus or minus one. We have already reported these results elsewhere.

In addition, in this report, the relation between the additional monopoles and Chiral symmetry breaking is discussed. We compute the Chiral condensate, the pseudo-scalar mass, and the pion decay constant. Preliminary results show that the additional monopoles do affect Chiral symmetry breaking. 
\end{abstract}

\maketitle


\section{Introduction}

  Monopoles are important topological QCD configurations, which are considered as responsible for colour confinement. The Kanazawa and Pisa groups have produced a number of Lattice results supporting dual superconductivity of QCD vacuum. Instantons relate to Chiral symmetry breaking as explained e.g. by the instanton liquid model by E. V. SHURYAK. However, a quantitative understanding of the relation between monopoles and instantons is rather difficult, also because monopoles are three dimensional objects, while instantons are four dimensional. We also expect that, if there is a relation between monopoles and instantons, monopoles would also affect Chiral symmetry breaking. 

To start we have done simulations to investigate the relations between instantons and monopoles. First, configurations are generated using the Wilson gauge action, and an Overlap Dirac operator is constructed from the gauge links. The eigenvalue problem of the Overlap Dirac operator is solved, and low-lying eigenvalues and eigenvectors are saved. We count the number of fermion zero modes in the configuration.

Next, we argue that we can calculate the number of instantons from the average square of the topological charge. In order to clarify quantitatively the relation between monopoles and instantons, we directly add one monopole and one anti-monopole with opposite charge to the configurations by the monopole creation operator. To check that the pair of monopoles is successfully added in the configurations, we look at the length of the monopole loops in the configurations.

Then, we count the number of zero modes by use of the Overlap fermions, and calculate the average square of the topological charge in the presence of the added monopoles. The result is that the addition of monopoles results in the addition of instantons. 

Moreover, we check that increasing the number of monopoles, being also a change in the number of instantons, affects Chiral symmetry breaking. We compute the Chiral condensate, the order parameter of Chiral symmetry breaking, the pseudo-scalar mass, and the pion decay constant. In the computations, the eigenvalues and eigenvectors that are computed from the normal configurations and the configurations with additional monopoles are used. The preliminary results show that the additional monopoles do affect Chiral symmetry breaking.

\section{Overlap fermions}

Wilson fermions are one of the most common formulations of quarks in Lattice gauge theory: a drawback is that they explicitly break Chiral symmetry. Therefore, we can not use them for this study. Instead of Wilson fermions, we use Overlap fermions. Overlap fermions preserve Chiral symmetry in Lattice gauge theory. The Ginsparg-Wilson relation~\cite{Gins1} describes Chiral symmetry in Lattice gauge theory, 
\begin{equation}
\gamma_{5} D + D \gamma_{5} = aDR\gamma_{5}D.
\end{equation}
$a$ is the lattice spacing, D is the Overlap Dirac operator, and R is a parameter. The right hand side of this relation is not zero, because of the Nielsen-Ninomiya no go theorem. By multiplying this Ginsparg-Wilson relation by the inverse of the Overlap Dirac operator $D^{-1}$ on both sides gives
\begin{equation}
\gamma_{5} D^{-1} + D^{-1} \gamma_{5} = aR\gamma_{5},
\end{equation}
showing that Chiral symmetry breaking of the propagator $D^{-1}$ is a local operator of $\mathcal{O}(a)$ vanishing in the continuum limit.

The form of the Overlap Dirac operator is defined in Ref.~\cite{Neuberger1} in terms of the massless Wilson Dirac operator $D_{W}$ (Wilson parameter: r = 1). 
\begin{equation}
D = \frac{1}{Ra} \left[ 1 + \frac{A}{\sqrt{A^{\dagger}A}}\right], \ A = - M_{0} + a D_{W}
\end{equation}
$M_{0}$ is a parameter, $0 < M_{0} < 2$. $D_{W}$ is the massless Wilson Dirac operator defined as follows:  
\begin{equation}
D_{W} = \frac{1}{2}[\gamma_{\mu}(\nabla_{\mu}^{*} + \nabla_{\mu}) - a\nabla_{\mu}^{*}\nabla_{\mu}]
\end{equation}
\begin{equation}
[\nabla_{\mu}\psi](n) = \frac{1}{a}[U_{n,\mu}\psi(n+\hat{\mu}) - \psi(n)]
\end{equation}
\begin{equation}
[\nabla_{\mu}^{*}\psi](n) = \frac{1}{a}[\psi(n) - U_{n-\hat{\mu}, \mu}^{\dagger}\psi(n-\hat{\mu})]
\end{equation}
In the numerical computations~\cite{Galletly1, Weinberg2}, the massless Overlap Dirac operator $D(\rho)$ is 
\begin{equation}
D(\rho) = \frac{\rho}{a} \left[ 1 + \frac{D_{W}(\rho)}{\sqrt{D_{W}(\rho)^{\dagger}D_{W}(\rho)}}\right].
\end{equation}
$D_{W}$ is computed from massless Wilson Dirac operator as follows:
\begin{equation}
D_{W}(\rho) = D_{W} - \frac{\rho}{a}, \ (\rho = 1.4). 
\end{equation}
$\rho$ is a (negative) mass parameter $0 < \rho < 2$. There are $n_{+}$ exact zero modes of plus chirality  and $n_{-}$ of minus chirality  in the spectrum of this massless Overlap Dirac operator. The topological charge is defined as $Q = n_{+} - n_{-}$, and the topological susceptibility $\langle Q^{2} \rangle$/V is computed from the topological charges.

The massless Overlap Dirac operator is calculated by the sign function using the Chebyshev polynomial approximation as follows:
\begin{align}
& \frac{D_{W}(\rho)}{\sqrt{D_{W}(\rho)^{\dagger}D_{W}(\rho)}} = sgn(D_{W}(\rho)) \equiv \gamma_{5}sgn(H_{W}(\rho))\\
& H_{W}(\rho) = \gamma_{5}D_{W}(\rho)
\end{align}
$H_{W}(\rho)$ is the Hermitian Wilson Dirac operator of $D_{W}(\rho)$. We use this $H_{W}(\rho)$ operator for computations of a minmax polynomial approximation~\cite{Giusti1}.

\subsection{Simulation details}

We generate configurations using the Wilson gauge action and periodic boundary conditions. In all our simulations, the number of iterations for the thermalization is $\mathcal{O}(2.0\times10^{4}$). Configurations are sampled after $\mathcal{O}(5.0\times10^{3}$) iterations between them. The numbers of configurations which we use in simulations are $\mathcal{O}(200) \sim \mathcal{O}(800)$ for each parameter $\beta$ and Volume, a total 17 sets of parameters. We construct the Overlap Dirac operator from gauge links of the configurations. We solve the eigenvalue problems $D(\rho)|\psi_{i}\rangle = \lambda_{i}|\psi_{i}\rangle$ using the subroutines ARPACK, and save $\mathcal{O}$(80) pairs of the low-lying eigenvalues $\lambda_{i}$ and eigenvectors $|\psi_{i}\rangle$. The index $i$ labels the pairs  ($1 \leqq i \leqq \mathcal{O}(80)$).

\subsection{Lattice spacing}

\begin{table}
\begin{tabular}{|c|c|c|c|c|c|c|c|} \hline
$\beta$ & $a^{(0)}$ [fm] & $a^{(1)}$ [fm] & $a^{(2)}$ [fm] & (n, $\alpha$) & Fit Range & $\chi^{2}/ndf$ & $N_{\mbox{conf.}}$\\ \hline
6.00 & $9.315\times10^{-2}$&  $9.2(4)\times10^{-2}$ & $9.0(7)\times10^{-2}$ & (20, 0.5, 5) & 2.0-6.0 & 1.19 & 440  \\ \hline
\end{tabular}
\caption{Determination of the lattice spacing. The lattice spacing $a^{(0)}$ is computed by the analytic function of~\cite{Necco1}. The lattice spacing $a^{(1)}$ and $a^{(2)}$ are computed by our simulations. n is the number of smearing steps and $\alpha$ is the weight factor of smearing.}\label{a1}
\end{table}

First, to fix the scale, we determine the lattice spacing by the analytic interpolation of Ref.~\cite{Necco1}. We check that by measuring the string tension and the $\bar q q$ potential from Wilson loops. We use APE smearing of the link variables to suppress excited states of the potential energy and fit the function $V(R) = \sigma R - \alpha/R + C$  to the potential $V(R)$. The results are listed in Table \ref{a1}.

As a check, from the values of the string tension $\sigma$ and of $\alpha$, the lattice spacing is computed in two different ways. (1) $a^{(1)}$: Sommer scale $r_{0} = 0.5 [\mbox{fm}]$, and (2) $a^{(2)}$: String tension $\sqrt{\sigma} = 440 [\mbox{MeV}]$ [Table \ref{a1}]. The results by simulations are consistent with the analytic interpolation. Therefore, we take the lattice spacing from the analytic interpolation~\cite{Necco1}, and use the Sommer scale $r_{0} = 0.5 \ [\mbox{fm}]$.

\subsection{Eigenvalues and Spectral density}

\begin{figure}[htbp]
 \includegraphics[width=55mm]{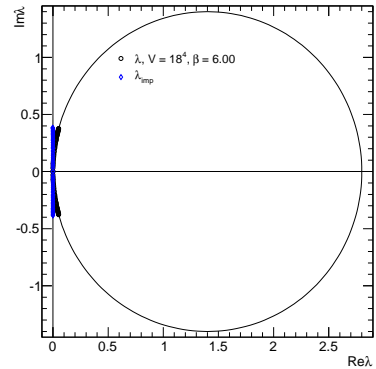}
 \caption{An example of distribution of eigenvalues $\lambda$, and improved eigenvalues $\lambda_{imp}$. There is one zero mode.}
 \label{fig:Dis_eigen1}
\end{figure}

\begin{figure}[htbp]
 \includegraphics[width=75mm]{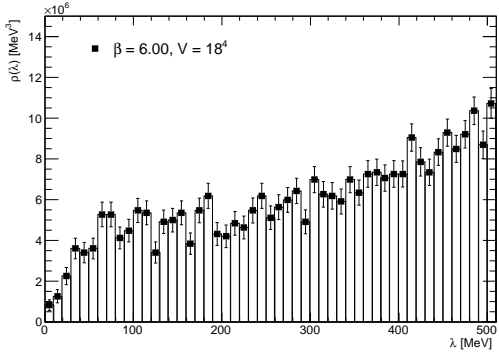}
 \caption{Spectral density of non zero modes ($\rho(\lambda)$). $V = 18^{4}$, and $\beta = 6.00$.}
 \label{fig:Spec_dens2}
\end{figure}

The eigenvalues $\lambda$ of the Overlap Dirac operator lie in the complex plane on a circle of center (1.4, 0) and radius 1.4, as shown in Figure \ref{fig:Dis_eigen1}, since we use a negative mass parameter $\rho = 1.4$. The eigenvalues $\lambda_{imp}$ of the improved massless Overlap Dirac operator $D^{imp}(\rho)$ are used~\cite{Capitani1}. They lie on the imaginary axis as shown in Figure \ref{fig:Dis_eigen1}. The improved massless Overlap Dirac operator $D^{imp}(\rho)$ is defined as:
\begin{equation}
D^{imp}(\rho) = \left( 1 - \frac{a}{2\rho} D(\rho) \right)^{-1} D(\rho)\label{imp_op1}
\end{equation}
The spectral density $\rho(\lambda, \ V)$ is defined as
\begin{equation}
\rho (\lambda, \ V) = \frac{1}{V}\langle \sum_{\lambda}\delta(\lambda - \bar{\lambda})\rangle.\label{eq:spec_1}
\end{equation}
 $\bar{\lambda} = \mbox{Im}{\lambda_{imp}}$. We show the spectral density $\rho (\lambda, \ V)$ except the zero modes in Figure \ref{fig:Spec_dens2}.

\subsection{The number of Zero modes, the topological charge, and the topological susceptibility}

\begin{figure}[htbp]
\includegraphics[width=75mm]{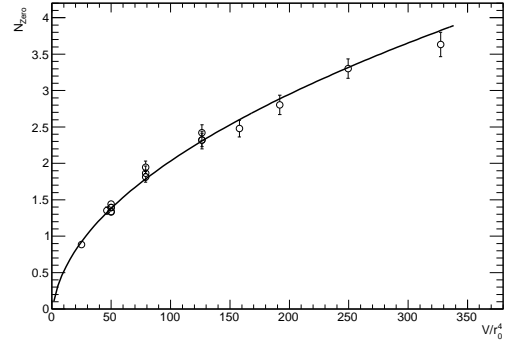}
\caption{The number of observed zero modes vs the physical volume. The continuous curve is $N_{Zero} = \sqrt{A*V/r_{0}^{4}} + B$, $A=4.9(3)$, $B = -0.18(5)$}
\label{fig:Zero_allV1}
\end{figure}

\begin{figure}[htbp]
\includegraphics[width=75mm]{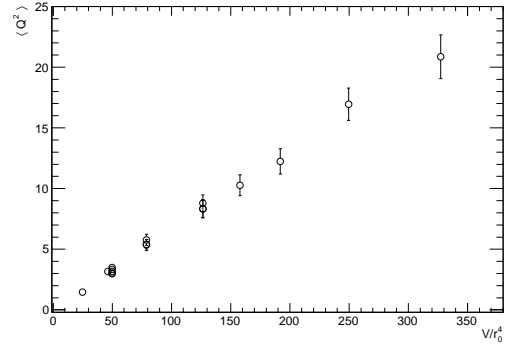}
\caption{The average square of the topological charges vs the physical volume. }
\label{fig:Q2_allv_2}
\end{figure}

In our simulations, we never observe zero modes of opposite chirality in the same configuration. The zero modes in our simulation have only + chirality or only - chirality in a given configuration. We suppose that the number of zero modes we observe is the \textbf{net} number of zero modes ($n_{+} - n_{-}$), that is to say the topological charges Q: in other words, we assume that pairs of zero modes of opposite chirality for some reason escape our detection. To check our supposition, we fit a function $N_{Zero} = \sqrt{A*V/r_{0}^{4}} + B$ to the number of zero modes as a function of $V$, Figure \ref{fig:Zero_allV1}. The result is $A = 4.9(3)\times10^{-2}$, $B = -0.18(5)$, and $\chi^{2}/ndf = 15.1/15.0$. The fitting range in the physical volume unit $V/r_{0}^{4}$ is from 24 to 330. Here, $N_{Zero}$ is the number of zero modes, averaged on the configurations. 

\begin{figure}[htbp]
  \includegraphics[width=75mm]{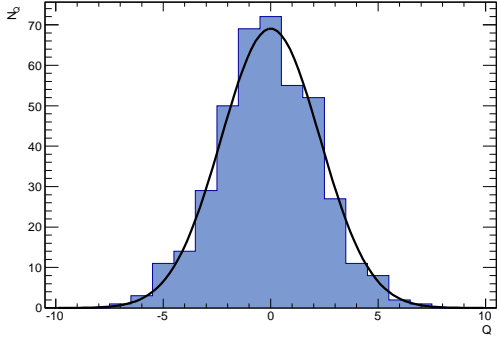}
\caption{A distribution of topological charges. The lattice is $V = 16^{4}$, and $\beta = 6.00$.}
\label{fig:Dis_topQ1}
\end{figure}

A typical distribution of the topological charges is that in Figure \ref{fig:Dis_topQ1} for ($V = 16^{4}$, $\beta = 6.00$): a Gaussian function $P_{Q} = \frac{e^{-\frac{Q^{2}}{2\langle Q^{2}\rangle}}}{\sqrt{2\pi\langle Q^{2}\rangle}}\{1 + \mathcal{O}(V^{-1})\}$~\cite{Giusti2} fits the data with $\chi^{2}/ndf = 6.9/13.0$, and the topological susceptibility $\langle Q^{2}\rangle r_{0}^{4}/V = 6.7(5)\times10^{-2}$. This value of topological susceptibility is consistent with a value $\langle Q^{2}\rangle r_{0}^{4}/V = 6.8(5)\times10^{-2}$ directly computed from the number of zero modes. Moreover, seventeen distributions of the topological charges are computed from our seventeen different lattices, and we fit the Gauss function to all distributions. All the resulting values of $\chi^{2}/ndf$ are in the range from 0.4 to 1.3. The topological charges have a Gaussian distribution.

Last, we fix the physical volume at $V/r_{0}^3 = 50.00$, and extrapolate the five data points of the topological susceptibility to the continuum limit using a linear expression $\langle Q^{2}\rangle r_{0}^{4}/V = c_{0} + c_{1} a^{2}$. We get for the topological susceptibility in the continuum limit: $(\chi  = 1.86 (6) \times 10^{2} \ \ [\mbox{MeV}])^{4} = 6.5 (5) \times 10^{-4} \ [\mbox{GeV}^{4}]$. We compare this result with Ref.~\cite{Witten1, Veneziano1, DelDebbio1, Giusti3}, and confirm that this result is consistent. Therefore, eigenvalues and eigenvectors of overlap fermions in our simulations are properly computed.

\section{Instantons}\label{sec:inst}

We would like to compute the number of instantons from the number of zero modes, but as anticipated, there are problems to determine it since some of them escape detection. In any translation invariant model indeed, e.g. the instanton liquid model, the number of instantons linearly increases with the physical volume. The number of our zero modes instead clearly increases as the square root of the physical volume [Figure \ref{fig:Zero_allV1}]. We never observe zero modes $n_{+}$ and $n_{-}$ of opposite chirality in the same configuration, namely the number of zero modes always coincides with the topological charge. Indeed the distributions of the topological charges determined in this way have Gaussian distributions with $ 0.4 \ < \ \chi^{2}/\mbox{ndf} \ < \ 1.3$, and agree with other groups results.

All that shows that we observe the \textbf{net} number of zero modes. At least at our rather small physical lattice volumes for some reason pairs of zero modes of opposite sign seem to appear as non-zero modes. This can explain why we obtain the correct topological charge anyway. To estimate the density of instantons we can use an analytic model based on the reasonable assumption that the instantons of both Chiralities are uniformly distributed in space-time and independent.

\subsection{The number of instantons}

Let us denote the number of instantons of positive chirality in a volume $V$  by $n_{+}$, the number of instantons of negative chirality by $n_{-}$. Of course 
\begin{equation}
\langle n_{+} \rangle =\langle n_{-} \rangle = \frac{N}{2} = \rho_{i} V,
\end{equation}
$\rho_{i}$ is the density of instantons. Because of $CP$ invariance instantons and anti-instantons have the same distribution. If the instantons are independent the distribution is Poisson-like.
\begin{equation}
\mbox{P}(n_{+}) = \frac{1}{n_{+}!}\left(\frac{N}{2}\right)^{n_{+}}\mathrm{e}^{\frac{-N}{2}} \label{n_{+}}
\end{equation}
\begin{equation}
\mbox{P}(n_{-}) = \frac{1}{n_{-}!}\left(\frac{N}{2}\right)^{n_{-}}\mathrm{e}^{\frac{-N}{2}} \label{n_{-}}
\end{equation}
The resulting distribution for $Q = n_{+} - n_{-} $ is
\begin{equation}
\mbox{P}(Q) = \sum^{\infty} _{n_{-}=0} P(n_{-}) P(n_{-}+Q) = \exp(-N) I_{Q}(N)
\end{equation}
Here $I_{Q} (x) $ Is the modified Bessel function 
\begin{equation}
I_{Q} (x) = \sum_{k=0}^{\infty} (\frac{x}{2})^{n_{-}} (\frac{x}{2})^{n_{-}+Q}\exp(-N) \frac{1}{n_{-}! (n_{-} +Q)!}
\end{equation}
When $Q\gg 1$, $N \gg1$ at $\frac{Q^2}{N}$ fixed 
\begin{equation}
\mbox{P}(Q) \simeq \frac{1}{\sqrt{2N\pi}}\mathrm{e}^{\frac{-Q^{2}}{2N}} \label{eq:pb_func1} 
\end{equation}
Finally, the number of instantons is determined as $N = \langle Q^{2} \rangle = \langle N_{Zero}^{2}\rangle$. 

\subsection{The instanton density}\label{sec:ins1}

\begin{figure}[htbp]
  \includegraphics[width=75mm]{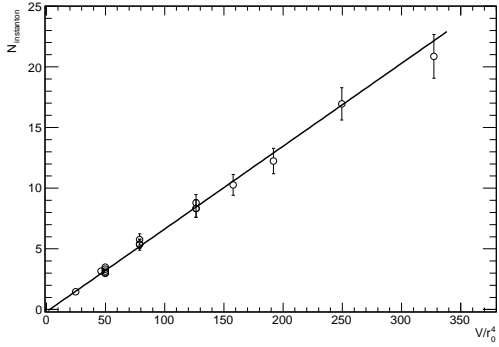}
\caption{The number of instantons ($\langle Q^{2} \rangle$) vs the physical volume.}
\label{fig:Q2_allV1}
\end{figure}

We obtain the instanton density by fitting a linear function $ N_{I} = 2\rho_{i}r_{0}^{4} * V/r_{0}^{4} + B$ to $\langle Q^{2} \rangle$ as shown in Figure \ref{fig:Q2_allV1}. The slope is $2\rho_{i}r_{0}^{4} = 6.8(2)\times 10^{-2}$ and the intercept is $B = -0.20(13)$. The intercept is compatible with zero.  Finally, the instanton density is 
\begin{equation}
\rho_{i} = 8.3 (3) \times 10^{-4} \ [\mbox{GeV}^{4}].
\end{equation}
This result is consistent with the instanton liquid model, Ref.~\cite{Shuryak1}.

\section{Monopoles}

In this section, we recall the definition of the monopole creation operator and the method to count monopoles. Our aim is to understand the relation between instantons and monopoles. To do that, we add monopole - antimonopole pairs of opposite charges in the configurations by the monopole creation operator. We then check whether the monopoles are successfully added in the configurations by counting the additional monopoles. 

\subsection{The monopole creation operator}

The monopole creation operator is defined in~\cite{DiGiacomo1, DiGiacomo2, DiGiacomo3}. Specifically, in this study, we use the monopole creation operator defined in~\cite{DiGiacomo3}, [Eq.(41) et seq.]. The monopole creation operator is defined as
\begin{equation}
\mu =\exp(-\beta \overline {\Delta S}) \label{m}
\end{equation}
$\overline {\Delta S}$ is defined by modifying the normal action at the time $t$ by replacing the usual plaquette $\Pi_{\mu\nu}(n)$ by $\bar{\Pi}_{\mu\nu}(n)$, as follows 
 \begin{equation} 
S + \overline{\Delta S} \equiv \sum_{n, \mu < \nu} \mbox{Re} (1 - \bar{\Pi}_{\mu\nu}(n))
\end{equation}
$\bar{\Pi}_{\mu\nu}(n) = \Pi_{\mu\nu}(n)$ in all sites $n$ with $n_0 \neq t$; at $n_0 = t$ the space-space components $i,j = 1 - 3$ are again unmodified $\bar{\Pi}_{i j}(t,\vec n) = \Pi_{i j}(t,\vec n)$ while $\bar{\Pi}_{i0}(t, \vec n)$ is as a modified plaquette with inserted matrices $M_{i}(\vec{n})$ and $M_{i}(\vec{n})^{\dagger}$, 
\begin{align}
& \bar{\Pi}_{i0}(t, \vec{n}) = \frac{1}{\mbox{Tr}\mbox[I]}\mbox{Tr}[U_ {i}(t, \vec{n})M_{i}^{\dagger}(\vec{n} + \hat{i})\nonumber \\
& \times U_{0}(t, \vec{n} + \hat{i})M_ {i}(\vec{n} + \hat{i})U_{i}^{\dagger}(t + 1, \vec{n})U_{0}^{\dagger}(t, \vec{n})]
\end{align}
$\mbox{Tr}\mbox[I]$ is the trace of the identity, and the matrix $M_{i}(\vec{n})$ is the discretised version of the classical field configuration $A_{i}^{0}(\vec{n} - \vec{x})$ produced by the monopoles to be added, namely 
\begin{equation} 
M_{i}(\vec{n}) = \exp(i A_{i}^{0}(\vec{n} - \vec{x})), \ ( i = x, \ y, \ z ).
\end{equation}

The form used for the monopole fields in a spherical coordinate system $(r, \ \theta, \ \phi )$ centred at the monopole is Wu-Yang:

\begin{align}
(i) \ n_{z} - z & \geqq 0  \nonumber \\
\begin{pmatrix}
A_{x}^{0} \\
A_{y}^{0} \\
A_{z}^{0}
\end{pmatrix}
& = 
\begin{pmatrix}
\frac{m_{c}}{2gr}\frac{\sin\phi (1 + \cos\theta)}{\sin\theta}\lambda_{3} \\
-\frac{m_{c}}{2gr}\frac{\cos\phi (1 + \cos\theta)}{\sin\theta}\lambda_{3} \\
0 
\end{pmatrix}
\end{align}

\begin{align}
(ii) \ n_{z} - z & < 0  \nonumber \\
\begin{pmatrix}
A_{x}^{0} \\
A_{y}^{0} \\
A_{z}^{0}
\end{pmatrix}
& = 
\begin{pmatrix}
-\frac{m_{c}}{2gr}\frac{\sin\phi (1 - \cos\theta)}{\sin\theta}\lambda_{3} \\
\frac{m_{c}}{2gr}\frac{\cos\phi (1 - \cos\theta)}{\sin\theta}\lambda_{3} \\
0 
\end{pmatrix}
\end{align}

The electric charge is
\begin{equation}
 g = \sqrt{\frac{6}{\beta}}: \ \mbox{gauge coupling constant}
\end{equation}

\noindent \textbf{We give the monopoles magnetic charges   $m_{c}= 0, 1, 2, 3, 4$ }		

One monopole has charge $+m_{c}$ and the other has charge $-m_{c}$. The total magnetic charge is zero. $m_{c}=0$ is the reference configuration with no monopoles added.

The monopole of $+m_{c}$ and anti-monopole of $-m_{c}$ are placed at time slice t, at a given spatial distance in the lattice. While Monte Carlo simulations are carried out, the pair of monopoles makes long monopole loops in the configurations. 

\subsection{The locations of the monopole and the anti-monopole}

\begin{figure}[htbp]
\includegraphics[width=55mm]{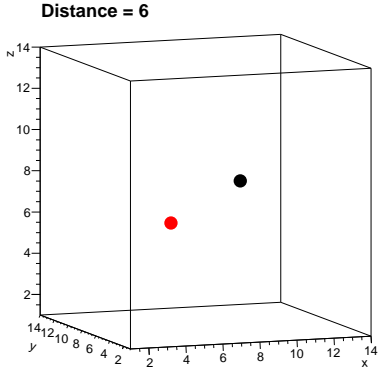}
\caption{The locations of the monopole and anti-monopole in x, y, z spaces. The lattice is $V = 14^{4} \ \beta = 6.00$.}
\label{fig:Box1}
\end{figure}

\begin{figure}[htbp]
\includegraphics[width=55mm]{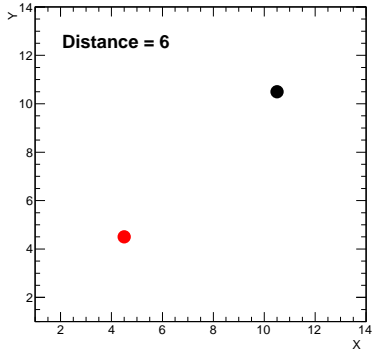}
\caption{The locations of the monopole and anti-monopole on x, y planes. The lattice is $V = 14^{4} \ \beta = 6.00$.}
\label{fig:Xy_plane1}
\end{figure}

We create the monopole-anti-monopole pair at time T = 7. The choice is irrelevant since boundary conditions are periodic in time. The locations of the monopoles in the lattice are chosen in the $(x,y)$ plane as shown in Figure \ref{fig:Box1} and Figure \ref{fig:Xy_plane1}. We define the distance between the monopole and anti-monopole as shown in Figure \ref{fig:Xy_plane1}. 

\subsection{Detecting the additional monopoles}\label{sec:ma_mono1}

To verify whether the monopoles are successfully added to the configurations, we detect the monopoles in the configurations~\cite{DeGrand1, Kanazawa1, Kanazawa2}. We generate the configurations varying the number of monopole charges from zero to four. The simulation parameters are listed in Table \ref{tb:length_conf_inf1}. The distance between the monopole and the anti-monopole is defined as Figure \ref{fig:Xy_plane1}.
\begin{table}[htb]
\begin{tabular}{|c|c|c|c|c|} \hline
$\beta$ & V & $N_{pairs}$ $\&$ $N_{charges} $ & Distance & $N_{Conf.}$ \\ \hline
6.00   & $14^{4}$ &   \mbox{Normal Conf.}   & - & 30 \\ \hline
   &     &   (1, 0)   &3&  30 \\
   &     &   (1, 1)   &4&  30 \\
 6.00   &  $14^{4}$  &   (1, 2)  &5 &  30 \\
  & &   (1, 3)   &7&  30  \\
  & &   (1, 4)   &7&  30  \\ \hline
\end{tabular}
\caption{The simulation parameters.}
\label{tb:length_conf_inf1}
\end{table}
Next, the configurations are iteratively transformed to the Maximally Abelian (MA) gauge using the Simulated Annealing algorithm. To remove the effects of the Gribov copies, 20 iterations are carried out in our simulations.

Abelian link variables, holding $U(1)\times U(1)$ symmetry, are derived by Abelian projection from non-Abelian link variables. The monopole current is defined for each colour direction as follows:  
\begin{equation}
k_{\mu}^{i} (n) = \frac{1}{2}\epsilon_{\mu\nu\rho\sigma}\partial_{\nu}n_{\rho\sigma}^{i}(n+\nu)
\end{equation}
The colour index can be i = 1, 2, and 3. $n_{\rho\sigma}^{i}(n+\nu)$ is the Dirac string~\cite{Poly1}, and the monopole current satisfies the conservation law,
\begin{equation}
\sum_{i}k_{\mu}^{i} (n) = \sum_{i}\frac{1}{2}\epsilon_{\mu\nu\rho\sigma}\partial_{\nu}n_{\rho\sigma}^{i}(n+\nu) = 0.
\end{equation}

\begin{figure}[htbp]
\includegraphics[width=75mm]{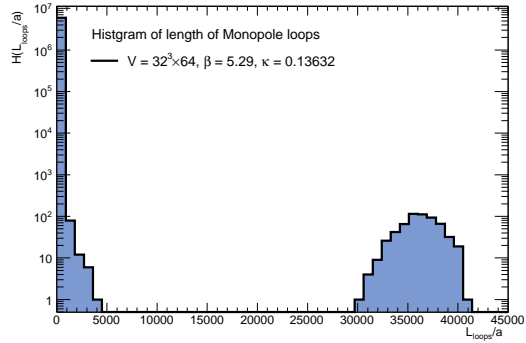}
\caption{The histogram of the length of the monopole loops. The normal configurations with two flavors of dynamical Wilson fermions, $V/r_{0}^{4} = 1.565(13)\times10^{3}, \ r_{0}/a = 6.05(5)$, are used.}
\label{fig:mon_clus2}
\end{figure}

The monopoles are known to form two clusters~\cite{Kanazawa2, Kanazawa3, Hart1} in MA gauge. The small (ultraviolet) clusters are composed of the short monopole loops. The large (infrared) clusters which percolate through the lattice and wrap around the boundaries of lattice are made of the longest monopole loop $L_{loops}$ in each color direction. The way how to compute numerically the monopole world line in four dimension is explained in~\cite{Bode1}. If the physical lattice volume is large enough, the small clusters and the large clusters are separated as in Figure \ref{fig:mon_clus2}. In quenched SU(2) study~\cite{Kanazawa3, Kanazawa4}, the long monopole loops only exist in confinement phase, and dominate the string tension. The long monopole loops are therefore considered to play an important role to produce color confinement. Going to the maximal abelian gauge is essential to divide monopole in clusters.

\begin{figure}[htbp]
\includegraphics[width=75mm]{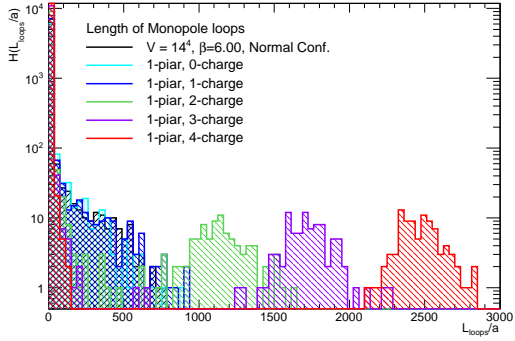}
\caption{The histogram of the length of the monopole loops. Monopoles with charges $m_c $ ranging from $0$ to $4$ are added to the configurations.}
\label{fig:Mon_loops_hist1}
\end{figure}

\begin{figure}[htbp]
\includegraphics[width=75mm]{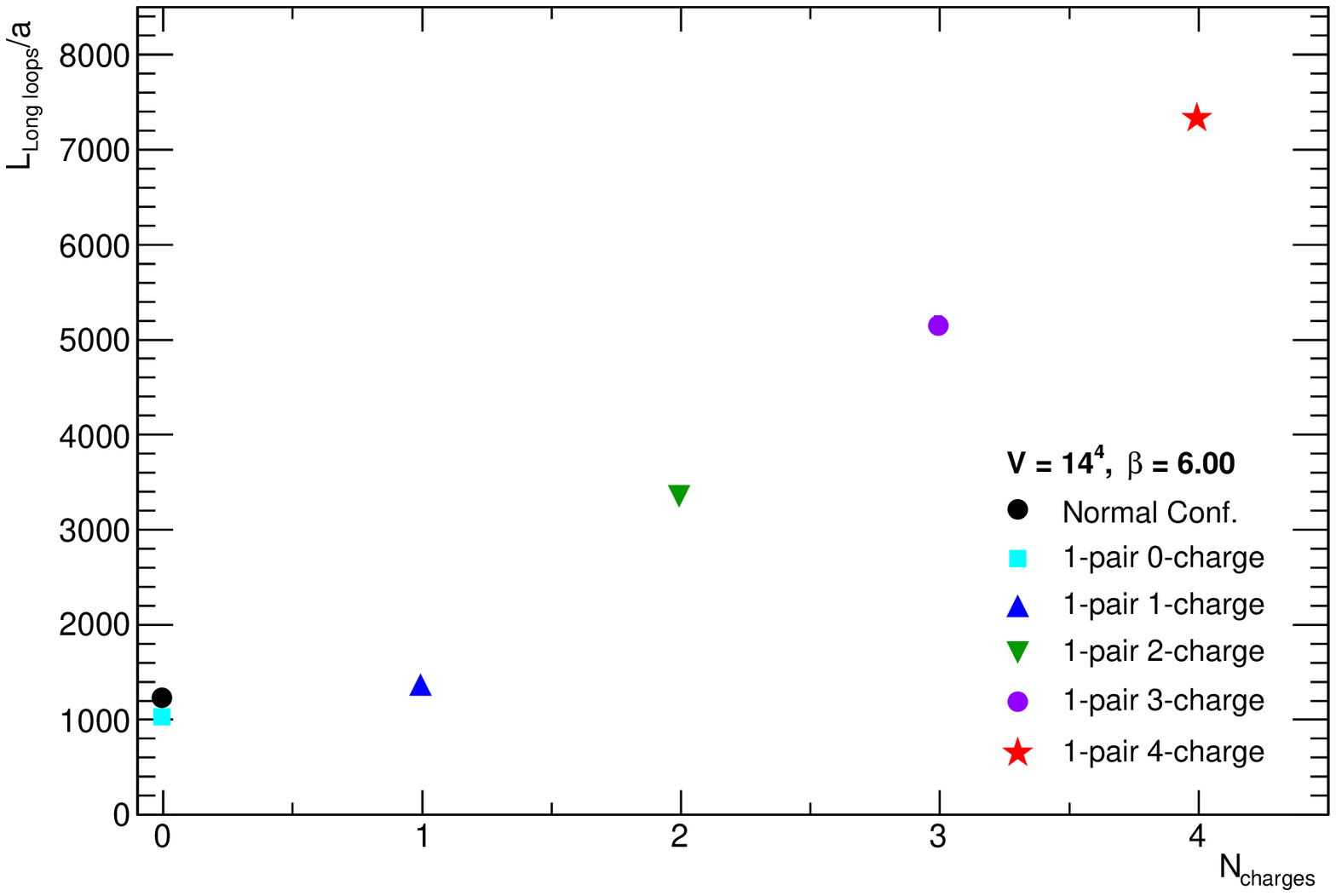}
\caption{The average of the long monopole loops divided by the number of configurations. The different colors and shapes of symbols indicate the different charges of monopoles.}
\label{fig:Mon_loops_long1}
\end{figure}

\begin{figure}[htbp]
\includegraphics[width=75mm]{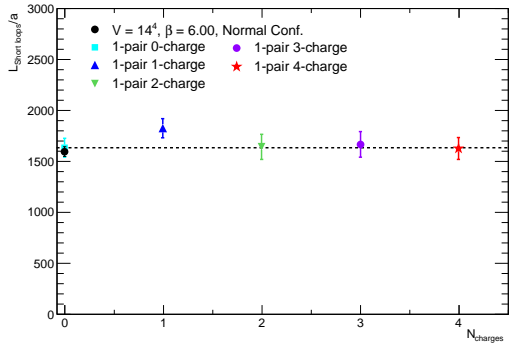}
\caption{The average of the short monopole loops divided by the number of configurations. The different colors and shapes of symbols indicate the different charges of monopoles.}
\label{fig:Mon_loops_short1}
\end{figure}

We create a histogram of the length of the monopole loops when one pair of monopoles with charges from zero to four are added as shown in Figure \ref{fig:Mon_loops_hist1}. To clarify which cluster increases with the monopole charges, we deduct the sum of the longest loop from the sum of all loops. We define the remainder of the subtraction as the sum of short loops. The averages of the sums of the long loops and short loops divided by the total number of configurations are computed respectively. The results are shown in Figure \ref{fig:Mon_loops_long1}, \ref{fig:Mon_loops_short1}. 

\section{The relations between Zero modes, instantons, and monopoles}

\subsection{Simulation details}

We generate configurations with one monopole-anti-monopole pair added with different magnetic charges. The distances between the monopole and anti-monopole are fixed at 6, and 8: slightly changing the distance between them, allows to check finite lattice volume effects. The simulation parameters are listed in Table \ref{tb:conf_inf1}. The Overlap Dirac operator is constructed from gauge links of the configurations. The eigenvalue problems are solved, and $\mathcal{O}$(60) pairs of the low-lying eigenvalues and eigenvectors are saved. We then count the number of zero modes, and calculate the average square of topological charges. Finally, we compare  the simulation results with an analytic prediction based on the hypothesis that the added monopoles do not perturb the distribution of instantons, but can only change the total topological charge.

Note: We do not do smearing, cooling, or MA gauge fixing in simulations. The number of zero modes, Distance 6: $N_{pairs}$ $\&$ $N_{charges} \ (1, 0)$, completely coincide with Distance 8. We take this as a check of volume independence.

\begin{table}[htb]
\begin{tabular}{|c|c|c|c|c|} \hline
$\beta$ & V & $N_{pairs}$ $\&$ $N_{charges} $ & Distance & $N_{Conf.}$ \\ \hline
6.00   & $14^{4}$ &   \mbox{Normal Conf.}   & - & 529 \\ \hline
   &  &   (1, 1)   &6&  312 \\
6.00  &  $14^{4}$ &   (1, 2)  &6 &  277 \\
   &  &  (1, 3)   &6&  259\\
  & &   (1, 4)   &6&  260  \\  \hline
   &  &   (1, 0)   &8&  400 \\
   &  &   (1, 1)   &8&  400 \\
6.00  &  $14^{4}$ &   (1, 2)  &8 &  410 \\
   &  &  (1, 3)   &8&  460 \\
  & &   (1, 4)   &8&  410 \\  \hline
\end{tabular}
\caption{The simulation parameters. }
\label{tb:conf_inf1}
\end{table}

\subsection{The number of zero modes and instantons}

\begin{figure}[htbp]
\includegraphics[width=75mm]{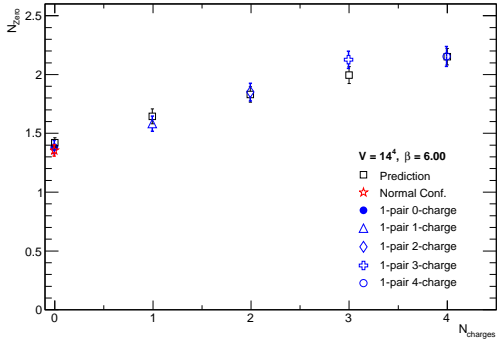}
\caption{The number of zero modes vs $N_{charges}= m_{c}$.}
\label{fig:Add_zero1}
\end{figure}

\begin{figure}[htbp]
\includegraphics[width=75mm]{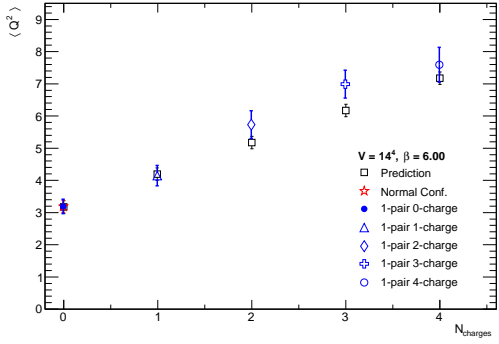}
\caption{The average square of topological charges vs $N_{charges}=m_{c}$.}
\label{fig:Add_q21}
\end{figure}

We plot the number of zero modes and the average square of the topological charges in Figure \ref{fig:Add_zero1}, and Figure \ref{fig:Add_q21} respectively as functions of the monopole charge, $N_{charges} =m_{c} $. $m_c =0$ is the ordinary case with no monopoles added.

These data are consistently explained by use of the following two arguments.

1) The creation operator Eq. (\ref{m}) acting on the vacuum produces a state with a pair of static monopoles propagating in time from $-\infty$ to $+\infty$. As a dual superconductor the vacuum shields them, so that they are one-dimensional structures and do not influence the external space time: $O(V^{-\frac{3}{4}})$ we expect that the distribution of zero modes Eq. (\ref{n_{+}}), \ (\ref{n_{-}}) is unaffected by the additional monopoles at large space-time volumes $V$.

2) A monopole-anti-monopole pair of charge $m_{c} =1$ produces one instanton or one anti-instanton with equal probability, as from $CP$ invariance. This correspondence one to one between monopole pairs and instantons has deep reasons, which will be discussed elsewhere, and our data will indeed support them. For larger values of $m_{c}$ the topology is the same as if we had $m_{c}$ monopole-anti-monopole independent pairs.

We recall that our detection only allows to detect instantons of the same chirality, i.e. the topological charge: pairs of instantons of opposite chirality escape detection.

From these assumptions we can predict $N_{Zero}$ and $\langle Q^2 \rangle$ analytically. The results are denoted as 'prediction' in Figures \ref{fig:Add_zero1} and \ref{fig:Add_q21}: they are also listed in Table\ref{tb:Add_zero_conc1}. To illustrate the procedure we present the case $m_{c} =2$ in some detail. We have two instantons with possible Chiralities $(+,+)$, $(-,-)$, $(+,-)$, $(-,+)$, which have equal probabilities, $\frac{1}{4}$. $\delta$ is the background number of instantons with distribution $P(\delta)$ Eq.(\ref{n_{+}}). In the case $(+,+)$ $\delta \rightarrow \delta +2$, for $(-,-)$  $\delta \rightarrow \delta -2$, for $(+,-)$, $(-,+)$ $\delta$ is unchanged.

Since we only can observe net values of chiralities we have
\begin{equation}
N_z = \frac{1}{4} \{ \langle |\delta +2| \rangle + \langle |\delta -2|\rangle\} +\frac{1}{2} \langle |\delta|  \rangle
\end{equation} 
which can be computed numerically using the probability function Eq.(\ref{n_{+}}). We also have
\begin{equation}
\langle Q^2 \rangle = \frac{1}{4} \{ \langle (\delta +2)^2 \rangle + \langle (\delta -2)^2\rangle\} +\frac{1}{2} \langle \delta^2  \rangle = \langle \delta^2  \rangle +2
\end{equation}
In the general case 
\begin{equation}
\langle Q^2 \rangle = \langle \delta^2  \rangle +m_{c}
\end{equation}
$\langle \delta^2  \rangle$ is nothing but $\langle Q^2 \rangle$ at $M_{c}=0$.

As a consistency check we have also computed on the configurations with no monopoles added the quantities $\langle |\delta \pm k| \rangle$ with $k$ an integer and from them $N_z $ and $\langle Q^2 \rangle$. The results are listed in Table \ref{tb:Add_zero_graph1} and agree, within errors, with the analytic determinations.
\begin{table}[htb]
\begin{tabular}{|c|c|c|c|} \hline
($N_{pair}$, $N_{charges}$) & $N_{|Q|}$ & $N_{z}$ & $N_{I}$ ($\langle Q^{2} \rangle$) \\ \hline
(1, 1)    & 1  & 1.64(6) & 4.2(2) \\
(1, 2)    & 2  & 1.83(6) & 5.2(2) \\
(1, 3)    & 3  & 2.00(7) & 6.2(2) \\
(1, 4)    & 4  & 2.15(7) & 7.2(2) \\ \hline
\end{tabular}
\caption{Prediction 1. $N_{|Q|}$ designates the number of the absolute value of the topological charges added by hands.}
\label{tb:Add_zero_conc1}
\end{table}

\begin{table}[htb]
\begin{tabular}{|c|c|c|c|} \hline
($N_{pair}$, $N_{charges}$) & $N_{|Q|}$ & $N_{z}$ & $N_{I}$ ($\langle Q^{2} \rangle$) \\ \hline
(1, 1)    & 1  & 1.59(6) & 4.2(2) \\
(1, 2)    & 2  & 1.78(6) & 5.1(3) \\
(1, 3)    & 3  & 1.96(7) & 6.2(4) \\
(1, 4)    & 4  & 2.12(7) & 7.2(4) \\ \hline 
\end{tabular}
\caption{Prediction 2. $N_{|Q|}$ designates the number of the absolute value of topological charges added by hands.}
\label{tb:Add_zero_graph1}
\end{table}

To quantify the consistency between prediction 1, 2, and the results by simulations, we fit a linear function $\langle Q^{2} \rangle = A * N_{charges} + B$ to Prediction 1, 2, and simulation results for $\langle Q^2 \rangle$ as a function of $m_{c}$. The final results are listed in Table \ref{tb:f_res1}. The results are consistent. Consequently, one monopole-antimonopole pair of one charge makes one instanton. 

\begin{table}[htb]
\begin{small}
\begin{tabular}{|c|c|c|c|c|} \hline
      & A & B & Fit Range ($N_{charges}$)      &$\chi^{2}/ndf$  \\ \hline 
  Prediction 1   &  1.00 (6)  &  3.2 (2)   & 1 - 5 & $\mathcal{O}(10^{-22})$/3.0 \\
  Prediction 2   &  1.10 (10) &  3.0 (3)   & 1 - 5 & 1.6/3.0 \\ \hline
  Distance 6     &  1.02 (13) &  2.90 (19) & 0 - 4 & 7.9/3.0 \\
  Distance 8     &  1.19 (11) &  3.1 (2)   & 0 - 4 & 1.4/3.0 \\ \hline
\end{tabular}
\caption{The final results.}\label{tb:f_res1}
\end{small}
\end{table}

\section{Monopoles and Chiral symmetry breaking}

In this section, we determine the fermion spectral density and from it the Chiral condensate $\langle\bar{\psi}\psi\rangle$, and $\langle\bar{\psi} \gamma_{5}\psi\rangle$. We also evaluate the pseudo-scalar mass $m_{ps}$, and from it the pion decay constant $f_{\pi}$ using the Gell-Mann-Oakes-Renner $GMOR$ relation~\cite{Gell-Mann1}.

\subsection{Simulation details}

In calculations, eigenvalues and eigenvectors computed from two different types of configurations are used: (1) Normal configurations. (2) Configurations with additional monopoles. The distance between the monopole and anti-monopole is 8. The simulation parameters are listed in Table \ref{tb:conf_inf1}.

First, we estimate the above physical quantities in the Chiral limit using eigenvalues and eigenvectors of normal configurations, varying the input quark mass in the range $ 0.5 \leqq \ m_{q} \ [\mbox{MeV}] \ \leqq \ 105.9 $. We show the results computed from normal configurations in Figures \ref{fig:psps_1}, \ref{fig:psps_2}, \ref{fig:psg5ps_1}, \ref{fig:mpi_1}, and \ref{fig:fpi_1}. In the figures, we use the black symbols to indicate the results which are computed from the normal configurations.

Then, we choose one input quark mass $m_{q} = 21.18 \ [\mbox{MeV}]$, and calculate the same physical quantities using the eigenvalues and eigenvectors computed from the additional monopole configurations. The results are shown in Figure \ref{fig:add_cond_1}, \ref{fig:add_cond_2}, \ref{fig:add_fv_1}, \ref{fig:add_mpi_1}, and \ref{fig:add_fpi_1}. We use the blue symbols to indicate the results which are calculated from  configurations with additional monopoles, and different  the  symbols for each value of the monopole charge.

A clear evidence that monopoles do affect chiral properties emerges from the comparison of the two sets of results. This time we do not consider the renormalization.

\subsection{Spectral density}
\begin{figure}[htbp]
    \includegraphics[width=75mm]{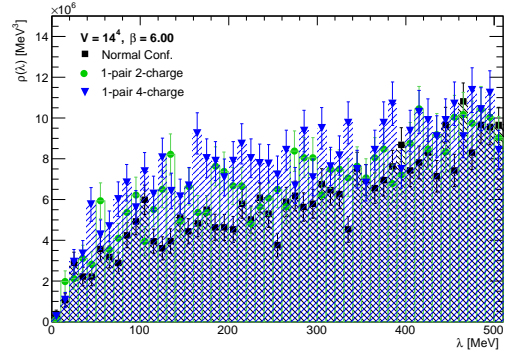}
    \caption{The spectral density of Overlap fermions.}
    \label{fig:spect_add2m2-4ch8d_1}
\end{figure}
The spectral densities except zero modes are computed by Eq. (\ref{eq:spec_1}), when the one pair of monopoles with two charges and four charges are added respectively. The results are plotted in Figure \ref{fig:spect_add2m2-4ch8d_1}.

\subsection{Chiral symmetry and flavor symmetry}

First, we define the fermion propagator $S_{F}$~\cite{Galletly1, Neff1}, 
\begin{equation}
S_{F} ( y - x )_{\alpha \beta, ab} \equiv \sum_{i}\frac{\psi_{\alpha ai}(x)\psi_{\beta bi}^{\dagger} (y)}{\lambda_{i}^{imp} + m_{q}}.
\end{equation}
The spinor indexes are $\alpha, \ \beta = 1, \ 2, \ 3, \ 4$, and color indexes are $a, \ b = 1, \ 2, \ 3$. The indexes $i, \ j$ denote the eigenvalue. $m_{q}$ is input quark mass. The normalization factor is $\sum_{x}\psi_{i}^{\dagger}(x)\psi_{i}(x) = 1.$ Therefore, we omit the normalization factor in our computations. The improved eigenvalues $\lambda_{imp}$ are computed from Eq. (\ref{imp_op1}). 

\begin{figure}[htbp]
  \includegraphics[width=75mm]{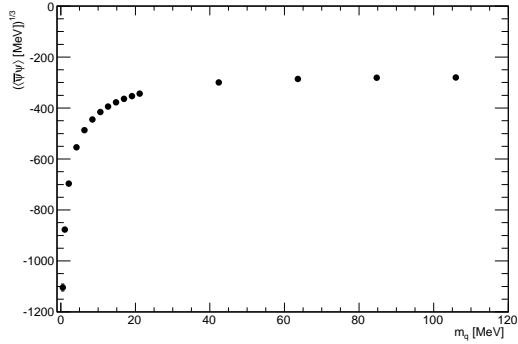}
\caption{The Chiral condensate in  physical units $(\langle \bar{\psi}\psi\rangle \ [\mbox{MeV}])^{1/3}$ vs the input quark mass. The normal configurations are used.}
\label{fig:psps_1}
\end{figure}

\begin{figure}[htbp]
  \includegraphics[width=75mm]{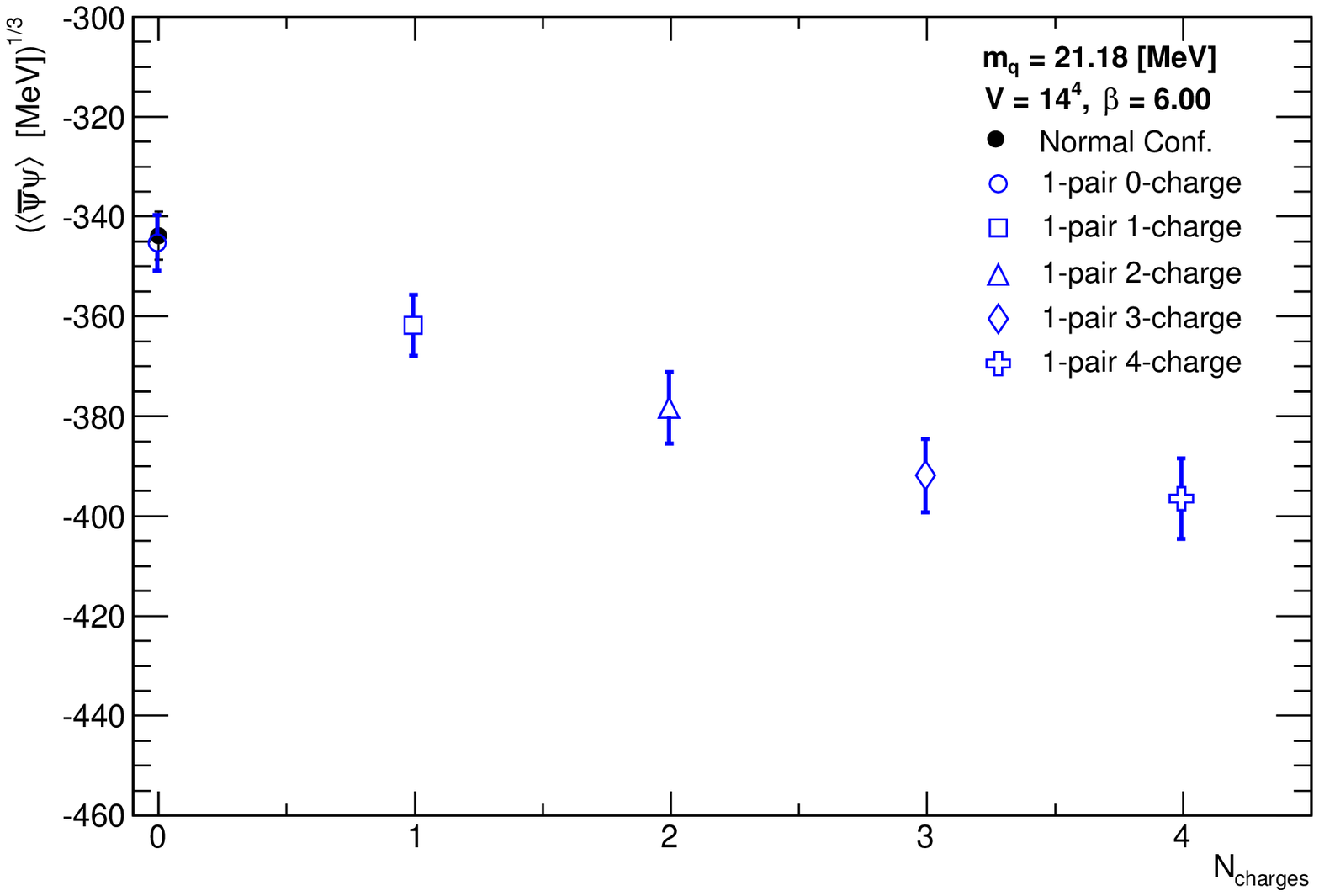}
\caption{The Chiral condensate in physical units $(\langle \bar{\psi}\psi\ \rangle[\mbox{MeV}])^{1/3}$ vs the monopole charge $N_{charges} = m_{c}$.}
\label{fig:add_cond_1}
\end{figure}

\begin{figure}[htbp]
  \includegraphics[width=75mm]{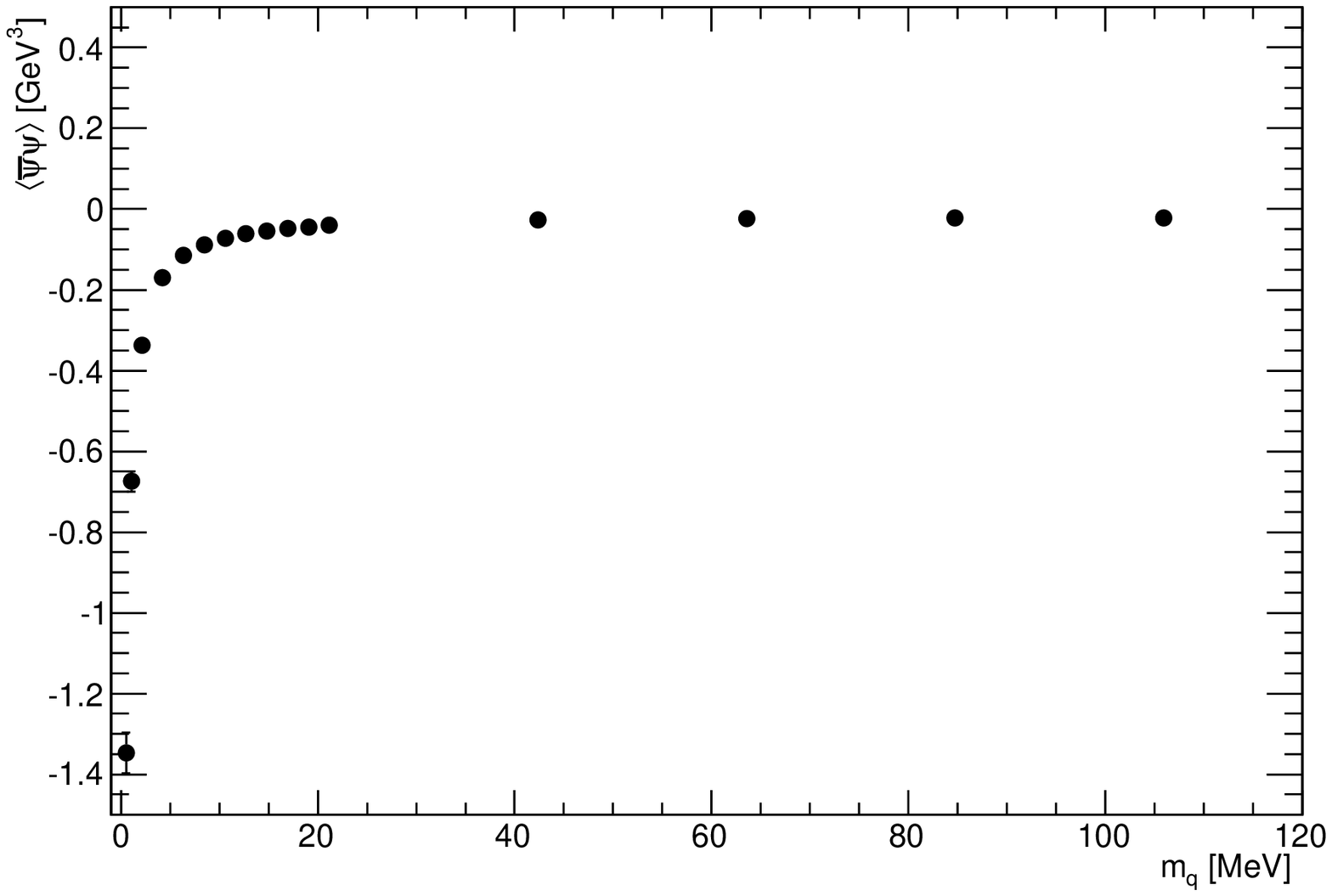}
\caption{The Chiral condensate in physical units $\langle\bar{\psi}\psi\rangle \ [\mbox{GeV}^{3}]$ vs the input quark mass. The normal configurations are used.}
\label{fig:psps_2}
\end{figure}

\begin{figure}[htbp]
  \includegraphics[width=75mm]{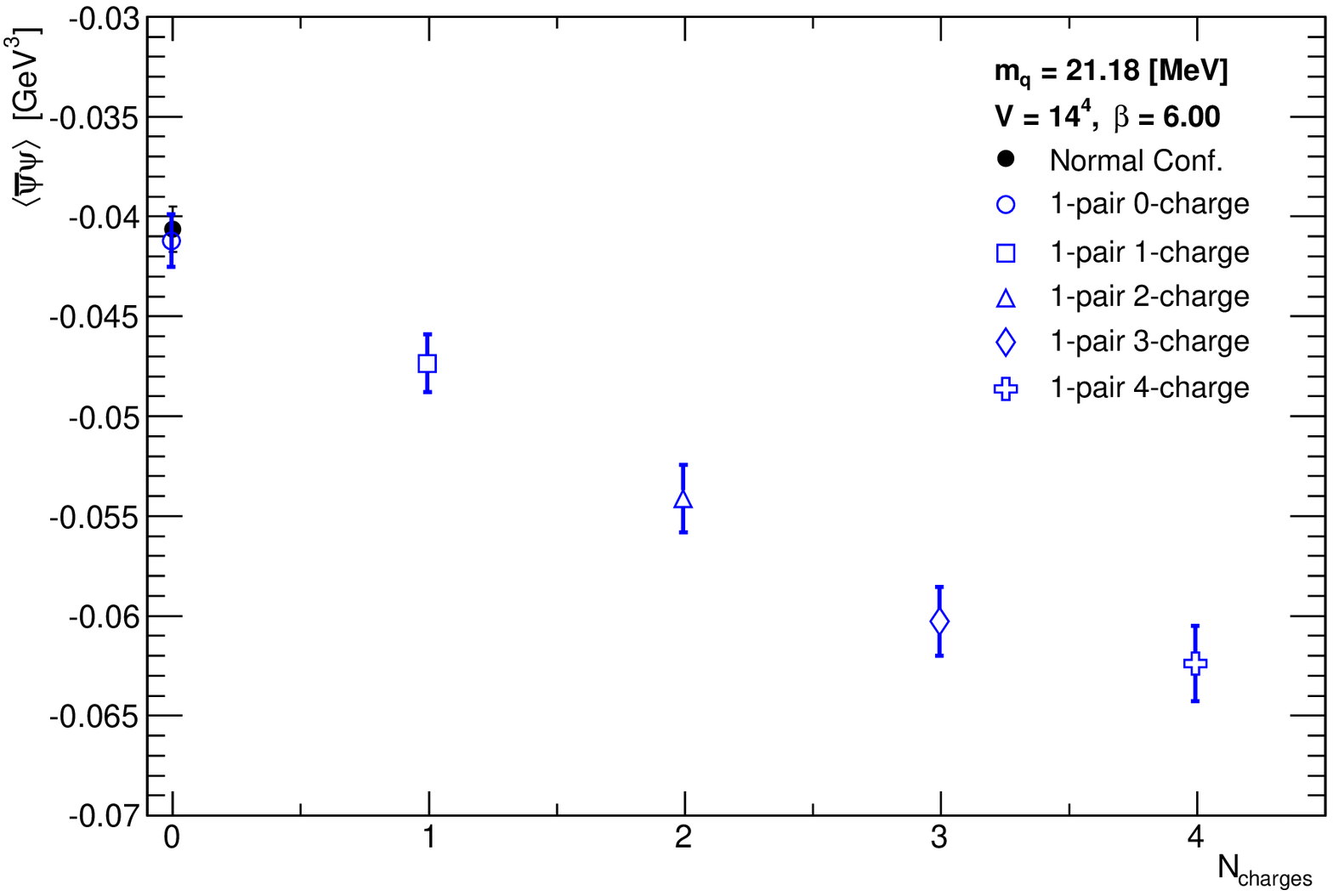}
\caption{The Chiral condensate in physical units $\langle\bar{\psi}\psi\rangle \ [\mbox{GeV}^{3}]$ vs the monopole charge $N_{charges} =m_{c}$.}
\label{fig:add_cond_2}
\end{figure}

The Chiral condensate that is an order parameter of Chiral symmetry breaks is computed by taking the sum of all of the indexes of the fermion propagator. 
\begin{align}
\langle \bar{\psi}\psi \rangle \equiv \langle \mathrm{tr} S_{F} (x - x)\rangle & = -\frac{1}{V}\sum_{x}\left( \sum_{i} \frac{\psi^{\dagger}_{i}(x)\psi_{i} (x)}{\lambda_{i}^{imp} + m_{q}}\right)\nonumber\\
& = - \frac{1}{V}\sum_{i}\frac{1}{\lambda_{i}^{imp} + m_{q}}
\end{align}
The Chiral condensates are computed in the two different physical units as shown in Figure \ref{fig:psps_1}, \ref{fig:psps_2}. We compare the Chiral condensates which are computed from the normal configurations and the additional monopole configurations as indicated in Figure \ref{fig:add_cond_1}, \ref{fig:add_cond_2}.

\begin{figure}[htbp]
  \includegraphics[width=75mm]{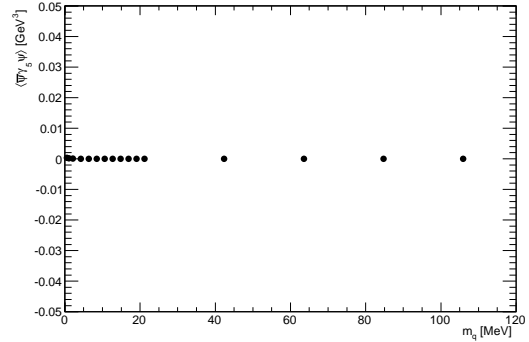}
\caption{The order parameter of the flavor symmetry $\langle\bar{{\psi}}\gamma_{5}\psi\rangle$ vs the input quark mass. The normal configurations are used.} 
\label{fig:psg5ps_1}
\end{figure}

\begin{figure}[htbp]
  \includegraphics[width=75mm]{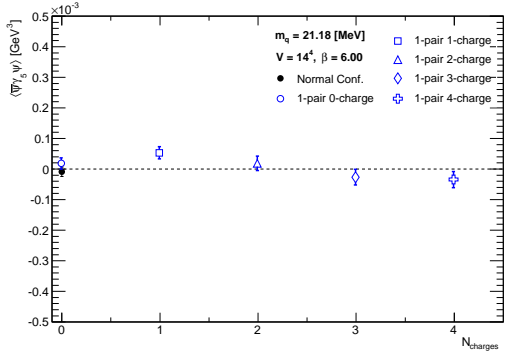}
\caption{The order parameter of the flavor symmetry $\langle \bar{{\psi}}\gamma_{5}\psi\rangle$ vs the monopole charge $N_{charges} = m_{c}$.} 
\label{fig:add_fv_1}
\end{figure}

Similarly, an order parameter of the flavor symmetry is defined as follows:
\begin{equation}
\langle \bar{\psi}\gamma_{5}\psi \rangle \equiv -\frac{1}{V}\sum_{x}\left( \sum_{i} \frac{\psi^{\dagger}_{i}(x)\gamma_{5}\psi_{i} (x)}{\lambda_{i}^{imp} + m_{q}}\right)
\end{equation}
The results are shown in Figure \ref{fig:psg5ps_1}, \ref{fig:add_fv_1}.

\subsection{The pseudo-scalar mass}

First, we compute two point functions from fermion propagators~\cite{Galletly1, DeGrand2}.
\begin{align}
&C_{\mbox{ps}} (\Delta t) \equiv \frac{1}{V}\sum_{t}\left[\sum_{ab}\sum_{ij}\sum_{\vec{x}}\sum_{\vec{y}}\frac{\left( \psi_{ai}^{\dagger} (\vec{x}, t) \gamma_{5} \psi_{bj} (\vec{x}, t) \right)}{\left( \lambda_{i}^{imp} + m_{q}\right)} \right. \nonumber \\
&\left. \times \frac{\left( \psi_{bj}^{\dagger} (\vec{y}, t + \Delta t) \gamma_{5} \psi_{ai}(\vec{y}, t + \Delta t) \right)}{\left(\lambda_{j}^{imp} + m_{q} \right)}\right]
\end{align}

\begin{figure}[htbp]
  \includegraphics[width=75mm]{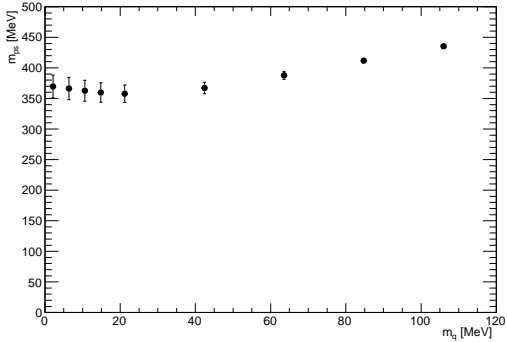}
\caption{$m_{\pi}$ vs the input quark mass. The normal configurations are used.} 
\label{fig:mpi_1}
\end{figure}

\begin{figure}[htbp]
  \includegraphics[width=75mm]{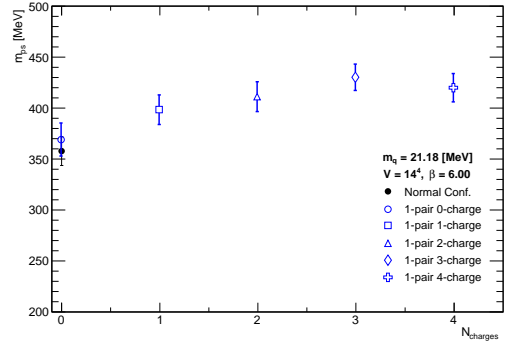}
\caption{$m_{\pi}$ vs the monopole charge $N_{charges} = m_{c}$.} 
\label{fig:add_mpi_1}
\end{figure}

The pseudo-scalar mass is estimated by fitting a function $f(t) = A_{0}\left[ \exp ( - m_{0} t) + \exp ( - m_{0} (L_{t} - t))\right]$ to the correlation between the pseudo-scalar and pseudo-scalar at the different time $\Delta t$. In this study, the fitting range is $2 \ \leqq \ \Delta t \ \leqq \ 12$. The results are shown in Figure \ref{fig:mpi_1}, \ref{fig:add_mpi_1}. The pseudo-scalar mass depends on the topological charges $|Q|$ Ref.~\cite{Galletly1}, and also, it increases with the monopole charges as shown in Figure \ref{fig:add_mpi_1}.

\subsection{Pion decay constant $f_{\pi}$}

\begin{figure}[htbp]
  \includegraphics[width=75mm]{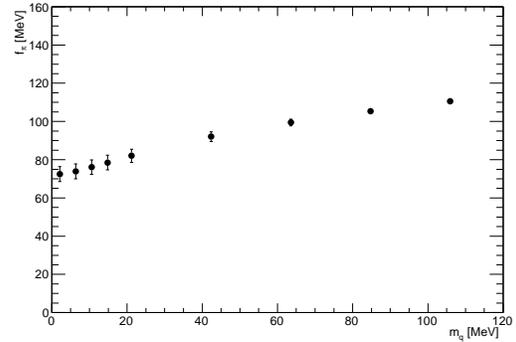}
\caption{The pion decay constant $f_{\pi}$ vs the input quark mass. The normal configurations are used.}
\label{fig:fpi_1}
\end{figure}

\begin{figure}[htbp]
  \includegraphics[width=75mm]{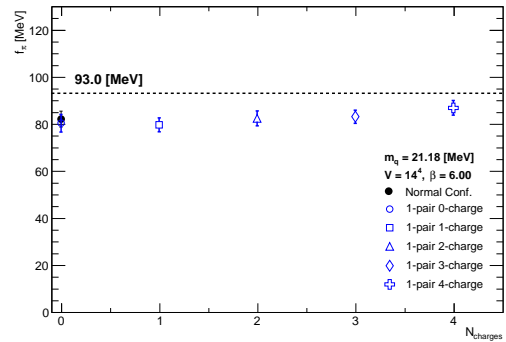}
\caption{The pion decay constant $f_{\pi}$ vs the monopole charge $N_{charges} = m_{c}$.}
\label{fig:add_fpi_1}
\end{figure}

According to the $GMOR$ relation~\cite{Gell-Mann1}
\begin{equation}
 \frac{(m_{u} + m_{d})}{2}\langle \bar{\psi}{\psi} \rangle = f_{\pi}^{2}m_{\pi}^{2},
\end{equation}
We can then determine the pion decay constant $f_{\pi}$ as
\begin{equation}
f_{\pi}  = \frac{\sqrt{m_{q}\langle \bar{\psi}\psi \rangle}}{m_{\pi}}, \ (m_{q} = m_{u} \sim m_{d}). 
\end{equation}
 The results are shown in Figure \ref{fig:fpi_1}, \ref{fig:add_fpi_1}.

\section{Summary}

We discussed the number of zero modes, the topological susceptibility in the continuum limit and the number of instantons, by adding monopoles into lattice configurations. We have studied the monopole loops in the configurations, and the relation between the number of zero modes and the charges of the monopoles. Preliminary results have been obtained on the effect of monopoles on Chiral symmetry breaking.

\section{Conclusion}

We have added a monopole and a anti-monopole to quenched $QCD$ configurations by use of a monopole creation operator.
\begin{itemize}
\item  The monopole and anti-monopole added form long monopole loops in a vacuum.
\item  Zero modes, that is instantons are created by the added monopole and anti-monopole.
\item The absolute value of the topological charges is increased.
\item Moreover, Chiral symmetry breaking is induced by the monopole - anti-monopole pair.
\end{itemize}
About the last sentence, we only have by now results for one value of the input quark mass and one lattice, and more work is needed. However, we already find some interesting indications 
\begin{itemize} 
\item Chiral condensate decreases by increasing the number of monopoles charges. 
\item The number of low-lying eigenvalues of the Overlap Dirac operator increases with the charge of the monopole.
\item The pion decay constant calculated from the Gell-Mann-Oakes-Renner relation remains unaffected by the monopoles.
\end{itemize} 
We keep running simulations in order to further clarify these relations. 


\begin{theacknowledgments}
We would like to thank sincerely E.-M. Ilgenfritz, Y. Nakamura, F. Pucci, G. Schierholz, and T. Sekido for useful discussions. This study is supported by the Research Executive Agency (REA) of the European Union under Grant Agreement No. PITN-GA-2009-238353 (ITN STRONGnet). M.H. received partial supports from I.N.F.N. at the University of Parma and the University of Pisa. We appreciate the computer resources and technical supports to the Research Center for Nuclear Physics and the Cybermedia Center at the University of Osaka, and the Yukawa Institute for Theoretical Physics the University of Kyoto. M. H. would like to thank the STRONGnet Summer School 2011 in Bielefeld for giving him the wonderful opportunity of discussing with F. Pucci about this study. And also, M. H. appreciates the welcoming cordialities from the University of Bielefeld, the University of Kanazawa, the University of Parma, and the University of Pisa.
\end{theacknowledgments}



\bibliographystyle{aipproc}   


\IfFileExists{\jobname.bbl}{}
 {\typeout{}
  \typeout{******************************************}
  \typeout{** Please run "bibtex \jobname" to optain}
  \typeout{** the bibliography and then re-run LaTeX}
  \typeout{** twice to fix the references!}
  \typeout{******************************************}
  \typeout{}
 }

\end{document}